%% file: main.tex
\documentclass[10pt]{article} 
\usepackage[nonatbib,preprint]{tmlr_modified}
\usepackage[numbers,sort&compress]{natbib}

\input{macros.tex}

\usepackage{url}

\title{Plan More, Debug Less: Applying Metacognitive Theory to AI-Assisted Programming Education\thanks{Preprint. Accepted as a paper at the AIED’25 conference.}}

\author{\name Tung Phung \email mphung@mpi-sws.org\\
        \addr Max Planck Institute for Software Systems
        \vspace{-2mm}
        \AND
        \name Heeryung Choi \email heeryung@umn.edu\\
        \addr University of Minnesota
        \vspace{-2mm}
        \AND
        \name Mengyan Wu \email mengyanw@umich.edu\\
        \addr University of Michigan 
        \vspace{-2mm}
        \AND      
        \name Adish Singla \email adishs@mpi-sws.org\\
        \addr Max Planck Institute for Software Systems
        \vspace{-2mm}
        \AND      
        \name Christopher Brooks \email brooksch@umich.edu\\
        \addr University of Michigan 
        \vspace{-2mm}
      }

\def\month{MM}  
\def\year{YYYY} 
\def\openreview{\url{https://openreview.net/forum?id=XXXX}} 

\begin{document}

\maketitle

\input{0_abstract}
\input{1_introduction}
\input{2_related_work}
\input{3_study_setup}
\input{4_RQ1}
\input{5_RQ2}

\input{6_limitations}
\input{7_conclusions}
\input{8_acknowledgements}

\bibliographystyle{unsrt}
\bibliography{main}

\end{document}

%% file: macros.tex
\usepackage[T1]{fontenc}
\usepackage{graphicx}
\usepackage{multirow}
\usepackage{subcaption}
\usepackage{xspace}
\usepackage{colortbl}
\usepackage[inline]{enumitem}
\usepackage{xcolor, soul}
\usepackage{arydshln}
\usepackage{setspace}
\usepackage{array}
\usepackage{makecell}
\usepackage{pifont}
\usepackage{diagbox}
\usepackage{hyperref}
\usepackage{marvosym}
\usepackage{wasysym}

\definecolor{planninghintcolor}{HTML}{91bdfa}
\definecolor{debugginghintcolor}{HTML}{e98281}
\definecolor{optimizinghintcolor}{HTML}{8fc287}
\definecolor{promptinputcolor}{HTML}{3725d9}
\definecolor{mygreen}{rgb}{0,0.6,0}
\definecolor{myred}{rgb}{0.9,0,0}

  {\list{}{\leftmargin=0.1in\rightmargin=0.0in}  \item[]  }%
  {\endlist}


%% file: 0_abstract.tex
\begin{abstract}
\looseness-1
The growing adoption of generative AI in education highlights the need to integrate established pedagogical principles into AI-assisted learning environments. This study investigates the potential of metacognitive theory to inform AI-assisted programming education through a hint system designed around the metacognitive phases of planning, monitoring, and evaluation. Upon request, the system can provide three types of AI-generated hints--planning, debugging, and optimization--to guide students at different stages of problem-solving. Through a study with 102 students in an introductory data science programming course, we find that students perceive and engage with planning hints most highly, whereas optimization hints are rarely requested. We observe a consistent association between requesting planning hints and achieving higher grades across question difficulty and student competency. However, when facing harder tasks, students seek additional debugging but not more planning support. These insights contribute to the growing field of AI-assisted programming education by providing empirical evidence on the importance of pedagogical principles in AI-assisted learning.

\end{abstract}

%% file: 1_introduction.tex
\section{Introduction} \label{sec:introduction}

\looseness-1Recent advancements in generative AI have sparked significant interest in the field of programming education~\cite{DBLP:journals/corr/abs-2402-01580,DBLP:conf/aied/MaCK24,DBLP:conf/icer/PhungPCGKMSS22,DBLP:conf/sigcse/WangMP24}, especially in the generation of personalized feedback~\cite{lohr2025you,DBLP:conf/lak/PhungPS0CGSS24,DBLP:conf/sigcse/WangMP24,xiao2024exploring}. However, existing studies often focus on technical correctness~\cite{gabbay2024combining} or student preference~\cite{DBLP:conf/aied/MaCK24,pankiewicz2024navigating} and overlook the importance of grounding AI-generated feedback in well-established pedagogical theories, potentially limiting the effectiveness of such feedback in student learning.

To address this gap, we propose enhancing AI-generated hints through the use of metacognitive scaffolds~\cite{loksa2022metacognition,Shin2023The,Volet1994Metacognitive}. Metacognitive scaffolds are instructional support mechanisms that help students plan, monitor, and evaluate their learning processes while fostering self-regulation and strengthening adaptive problem-solving skills~\cite{flavell1979metacognition,holton2006scaffolding,schraw1995metacognitive}. These scaffolds are crucial in programming education, where students often struggle with structuring approaches to solving (planning)~\cite{ebrahimi2006taxonomy,parsons2023exploring}, identifying and fixing errors (monitoring)~\cite{park2025exploring}, and optimizing solutions (evaluation)~\cite{DBLP:conf/sigcse/SalibaSOCQ24}. By grounding AI-generated hints in these metacognitive phases, we aim to provide not only technical assistance but also structured yet flexible support to promote students' metacognitive development. 
Specifically, we design three corresponding types of hint: \emph{planning}, \emph{debugging}, and \emph{optimization}, which we collectively refer to as \textbf{AI}-generated hints based on \textbf{M}etacognitive \textbf{S}caffolds (AIMS hints). To foster students' metacognitive awareness of their problem-solving stages, we adopt a learning-assisted approach in which we set a quota for total hints per question and let the students decide for themselves which hint type to request based on their needs. 
Figure~\ref{fig:illustration_help_seeking} provides an overview of how students utilized these hints in our deployment of this system.

Our study is centered on the following research questions:

\begin{itemize}[leftmargin=15pt,itemsep=-3pt,topsep=-2pt,label={}]
    \item \textbf{RQ1:} How do AIMS hints impact students' help-seeking behaviors?
    \item \textbf{RQ2:} How do those behaviors relate to students' problem-solving performance?
\end{itemize}

\input{figs/illustrative_patterns/fig_main}

We examine these questions across all students and subsets based on question difficulty and student competency levels. Our contributions are as follows:
\begin{itemize}[leftmargin=15pt,itemsep=-3pt,topsep=-2pt]
    \item[$\bullet$] \textbf{Hint and system design.} Using metacognitive theory, we design a hint system that offers three types of AIMS hints: planning, debugging, and optimization, and evaluate these hints through a classroom field study.
    \item[$\bullet$] \looseness-1\textbf{Student behavior analysis.} We analyze student help-seeking behaviors, revealing trends such as students value planning hints highly but often underutilize them in favor of debugging hints, especially when facing harder tasks.
    \item[$\bullet$] \textbf{Student performance analysis.} We find that planning hints are linked to better performance, notably for higher-competency students.
    \item[$\bullet$] \textbf{Code release.} To enhance reproducibility and aid future research, we publicly release the implementation of our AIMS hint generation techniques.
\end{itemize}

By integrating metacognitive scaffolds into AI-generated hints, this work not only contributes to a deeper understanding of how personalized AI support can enhance programming education but also provides practical insights for designing pedagogically-grounded educational AI systems.

%% file: figs/illustrative_patterns/fig_main.tex
\begin{figure}[t!]
        \includegraphics[width=\linewidth]{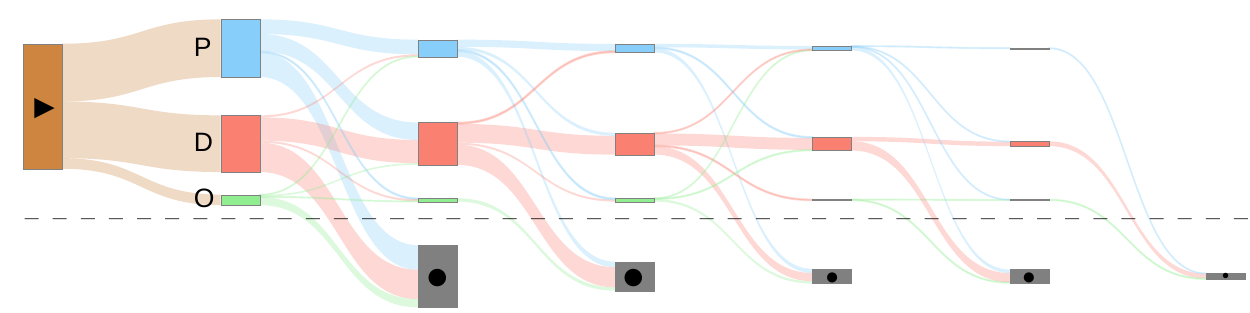}
    \vspace{-7mm}
    \caption{
        Overview of requested hints (725 hints in 366 student-question pairs). Throughout the paper, P, D, and O denote a planning, debugging, and optimization hint, respectively. \Forward{} marks the beginning and $\CIRCLE{}$ marks the end of problem-solving. The graph displays sequences of hints, link sizes depicting counts. 
    }
    \label{fig:illustration_help_seeking}
    \vspace{-4mm}
\end{figure}

%% file: 2_related_work.tex
\section{Related Work}
\textbf{Metacognitive scaffolds and self-regulated learning.}
Many efforts have been made to incorporate metacognitive scaffolds in programming education. The benefits of metacognition on learning behaviors and performance have been consistently shown through research \cite{Choi2023-fg,2017Metacognitive,Shin2023The,Vieira2019Student,Volet1994Metacognitive,Yilmaz2021Learning}. For example, Vieira et al.~\cite{Vieira2019Student} found that novice computer science students wrote longer self-explanation in-code comments compared to experienced peers because they saw self-explaining as a learning opportunity. Choi et al.~\cite{Choi2023-fg} showed that prompting reflection after programming tasks was correlated with better learning perception and performance on both immediate and delayed post-tests. Yilmaz and Yilmaz \cite{Yilmaz2021Learning} showed that students who received personalized metacognitive feedback weekly engaged significantly more in a Computer I course. Inspired by these insights, we ground in metacognitive theory to design AI-generated scaffolding hints.

\looseness-1\textbf{AI-generated feedback in programming education.} 
Recent research has explored AI-generated feedback in programming education~\cite{DBLP:conf/sigcse/WangMP24,neurips2023gaied_32_zamfirescu-pereira}. 
Phung et al.~\cite{DBLP:conf/lak/PhungPS0CGSS24} found that providing symbolic information such as the buggy output and fixed programs in prompts can improve the quality of AI-generated debugging hints.
Lohr et al.~\cite{lohr2025you} showed that AI can be directed to provide feedback that focuses on specific aspects, such as knowledge about task constraints or performance. 
Xiao et al.~\cite{xiao2024exploring} explored students' perceptions of hints with varying detail levels, revealing that their effectiveness depends on context and that high-quality next-step and debugging hints do not always facilitate student progress.
Building on this line of work, our study employs AI techniques that utilize symbolic information to generate different types of hints tailored to students' needs.
Our hint system utilizes a button-based interface, as opposed to a chat-based one~\cite{DBLP:conf/aied/MaCK24,sun2024investigating}, since the latter's pedagogical effects are still unclear.

\looseness-1\textbf{Students' help-seeking behaviors with automated hints}. Several studies have examined how students interact with automated hints~\cite{li2023college,wiggins2021exploring}. Marwan et al.~\cite{marwan2019evaluation} found that data-driven next-step hints improved immediate performance, and when paired with self-explanation prompts, led to learning gains. Expanding on this, we investigate the association between AI-generated hints and student performance. Price et al.~\cite{price2017hint} found that the quality of initial hints positively correlates with help abuse. To address this, we set a fixed quota for hint use to prevent over-reliance on help. Wiggins et al.~\cite{wiggins2021exploring} characterized student's hint-seeking behavior along two axes of elapsed time and code completeness, while Bui et al.~\cite{bui2024hints} explored different hint formats, including text and skeleton code. Our study extends this research by analyzing additional aspects, including engagement, perception, and hint request sequence in the context of AIMS hints.

\textbf{AI support and metacognition.} Recently, concerns have arisen that AI technologies might reduce students' engagement in metacognitive practices, a concept referred to as \emph{Metacognitive laziness}~\cite{Fan2024-hv}. Our work explicitly addresses this by integrating metacognitive theory into AI-generated hints, aiming to support self-regulated learning (SRL) skills development rather than replace it.\\

%% file: 3_study_setup.tex
\section{Study Setup}

\looseness-1This section outlines the study context, our proposed AIMS hints, the deployment, and methods to estimate question difficulty and student competency.

\subsection{Course and Students}  \label{sec:study_setup.course}

\looseness-1\textbf{Course overview.}
This study was conducted in a Python-based introductory data science course as part of an online Master's program at the University of Michigan. The four-week course featured weekly assignments covering key topics such as regular expressions, \emph{pandas} data frame manipulation, Excel processing, and CSV file handling. Each assignment, delivered as a Jupyter notebook, consisted of three to four programming questions ($14$ in total), requiring students to complete Python functions for specific tasks. Assignments were due weekly, and students could submit multiple times before the deadline, with their highest score counting toward the final grade.

\looseness-1\textbf{Student overview.}
Overall, $102$ students enrolled in the course: $71$ males, $27$ females, and four unspecified. Their ages range from 18 to 58 (mean = $32$, stddev = $8.5$).
Requesting hints was voluntary, with no additional incentives or penalties, and instructors were unaware of whether or how often students requested hints. Students were informed about the research aspect of the initiative, including anonymous data recording and the AI-generated nature of hints which might not always be correct. This study was deemed exempt from oversight by the Institutional Review Board under application number HUM00251143.

\subsection{Hint Types and AI-Generation Techniques}  \label{sec:study_setup.hint_techniques}

\looseness-1\textbf{AIMS hints.}
We mapped the three metacognitive phases of planning, monitoring, and evaluation onto disciplinary terms of planning, debugging, and optimization. Each hint type was designed with a specific goal: planning hints assist in initial strategy formulation, debugging hints aid in issue identification and resolution, and optimization hints foster code quality reflection and improvement for students aiming to exceed assignment expectations (see Figure~\ref{fig:hint_types}).

\input{figs/hint_types/fig_main}

\looseness-1\textbf{Techniques for generating AIMS hints.}
Each hint type is generated by a technique, all with careful consideration of incorporating “guard-rails” instructions~\cite{liffiton2023codehelp} to prevent the AI model (GPT-4o~\cite{hurst2024gpt}) from revealing the solutions.\footnote{\url{https://github.com/machine-teaching-group/aied2025-plan-more-debug-less}} Our technique for debugging hints is adapted from previous studies that showed good performance in data science programming education~\cite{DBLP:conf/lak/PhungPS0CGSS24,neurips2023gaied_32_zamfirescu-pereira}. It follows a two-phase process: First, it extracts symbolic information including (1) the buggy output (obtained from running the student's program) and (2) a repaired program (obtained from requesting the AI model). Second, it uses this information along with the student's code and any reflection (detailed in Section~\ref{sec:study_setup.hint_system}) to prompt the AI to generate an explanation (leveraging Chain-of-Thought~\cite{wei2022chain}) and a Socratic-style hint for a single bug (to be provided to the student).
Since there were no existing techniques for generating high-quality planning and optimization hints, we adjust this technique for the other two hint types. 
For planning hints, we modify the prompt's language to focus on problem-solving steps rather than debugging and remove the repaired program to shift emphasis away from errors.
For optimization hints, we follow a similar two-phase process but replace the repaired program with an AI-generated optimized program--focusing on short running time while still requiring correctness.

\subsection{Hint System and Student Interaction}  \label{sec:study_setup.hint_system}
\looseness-1\textbf{AIMS hint system.} We develop a hint system consisting of two main components: (1) a backend server for generating hints using the techniques introduced in Section~\ref{sec:study_setup.hint_techniques} and (2) a JupyterLab extension as the interface for interacting with students.
To prevent over-reliance on AI and foster students' metacognitive awareness, we set a quota limit of five hints per question and allow students to choose the type of hint to request. The extension displays three buttons below each assignment question, enabling students to request planning, debugging, or optimization hints. Before the course, the instructor introduced the hint system and demonstrated how to use it to the class. Students could always click a "?" button located next to the hint buttons to view descriptions of the hint types (as in Figure~\ref{fig:hint_types}). The first time a student requests a hint, a “Consent” pop-up informs them about the research aspect of the system (see Section~\ref{sec:study_setup.course}). Students can only proceed to request hints after agreeing to this notice.
To ensure easy access to previous hints, each hint is stored in a collapsible widget below the corresponding question's hint buttons. When a student reopens a notebook, these widgets are collapsed by default and can be expanded with a click to revisit previous hints.

\looseness-1\textbf{Student interaction.} Figure~\ref{fig:hint_request_interaction} demonstrates the interaction between students and our system for requesting hints. When requesting hints, students are encouraged to reflect on their progress or issues. These reflections serve dual purposes: promoting students' engagement by prompting them to articulate their thoughts and providing the AI model with context to generate a relevant hint. Once the backend generates a hint, it is sent back to the interface and displayed to the student, accompanied by two feedback buttons: “thumb up” and “thumb down”, allowing the student to rate the hint as helpful or unhelpful, respectively.

\subsection{Data Collection}  \label{sec:study_setup.data_collection}

\looseness-1We collected comprehensive data on student behavior and performance, including hint requests and revisits, assignment submissions, and final solving status. Of $102$ students, $101$ ($99\%$) activated the JupyterLab extension, and $76$ ($75\%$) requested at least one hint. Figure~\ref{fig:illustration_hint_types} shows examples of provided hints. Figure~\ref{fig:basic_stats} provides a breakdown of students and hints across $14$ assignment questions.

\input{figs/hint_request_interaction/fig_main}

\input{figs/illustrative_hint_types/fig_main}

\input{figs/basic_stats/fig_main}

\subsection{Question Difficulty and Student Competency}  \label{sec:study_setup.difficulty_competency}
To analyze the impacts of AIMS hints across varying conditions, we categorize questions by difficulty and students by competency.
Question difficulty is estimated using past student performance in two prior iterations of the same course: the easier (harder) group consists of four highest- (lowest-) scored questions, one per assignment.
Student competency is approximated based on the number of attempts until solving all questions in Assignment 1: fewer attempts indicate higher competency.
We designate the top third (34 students) as the higher-competency group and the bottom third (34 students) as the lower-competency group. Since Assignment 1 served as a proxy, it is excluded from competency-based analyses.

%% file: figs/hint_types/fig_main.tex
\begin{figure}[t!]
    \scalebox{0.90}{
        \setlength\tabcolsep{5pt}
        \renewcommand{\arraystretch}{1.5}
        \input{figs/hint_types/content}

    }
    \vspace{-2mm}
    \caption{
        AIMS hint types with the descriptions provided to students.
    }
    \label{fig:hint_types}
    \vspace{-3mm}
\end{figure}

%% file: figs/hint_types/content.tex
\begin{tabular}{cc}
    \hline
    \textbf{Hint type} & \textbf{Description} \\
    \hline
    \multicolumn{1}{p{0.14\linewidth}}{\multirow{1}{*}{\colorbox{planninghintcolor}{\textcolor{white}{\phantom{:::}Planning\phantom{||--}}}}}
    & \multicolumn{1}{p{0.92\linewidth}}{A hint aimed at helping you to identify the steps needed to solve the question.} \\
    \hline
    \multicolumn{1}{p{0.14\linewidth}}{\multirow{1}{*}{\colorbox{debugginghintcolor}{\textcolor{white}{\phantom{::}Debugging\phantom{::}}}}}
    & \multicolumn{1}{p{0.92\linewidth}}{A hint aimed at helping you to identify and fix a bug in your current program.} \\
    \hline
    \multicolumn{1}{p{0.14\linewidth}}{\multirow{1}{*}{\colorbox{optimizinghintcolor}{\textcolor{white}{Optimization}}}}
    & \multicolumn{1}{p{0.92\linewidth}}{A hint aimed at helping you to optimize your current program for better performance and readability.} \\
    \hline
\end{tabular}

%% file: figs/hint_request_interaction/fig_main.tex
\begin{figure}[t!]
    \centering
    \includegraphics[width=0.95\linewidth]{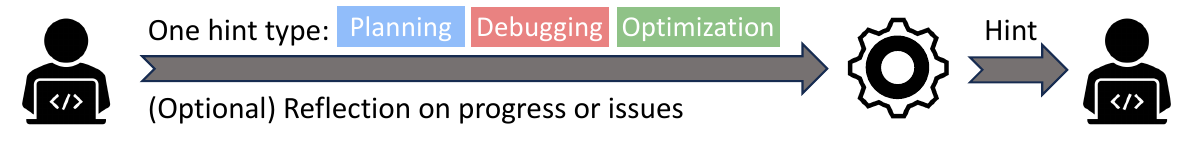}
    \vspace{-3mm}
    \caption{
         Interaction between a student and our hint system for requesting a hint.
    }
    \label{fig:hint_request_interaction}
    \vspace{-2mm}
\end{figure}

%% file: figs/illustrative_hint_types/fig_main.tex
\begin{figure}[t!]
    %
    \includegraphics[width=\linewidth]{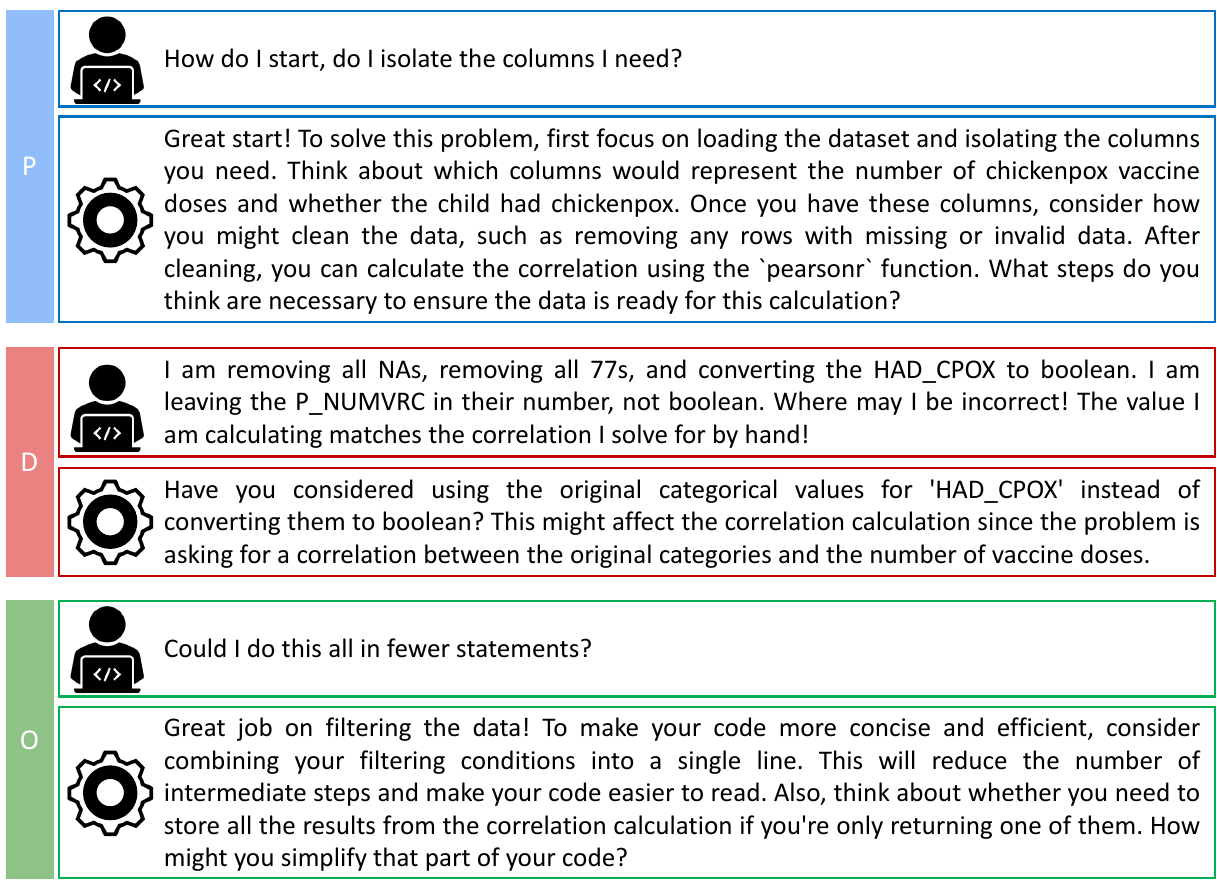}
    \vspace{-7mm}
    \caption{
         Examples of hint types: student reflections and received hints in the study.
    }
    \label{fig:illustration_hint_types}
    \vspace{-2mm}
\end{figure}

%% file: figs/basic_stats/fig_main.tex
\begin{figure}[t]
    %
    \includegraphics[width=\linewidth]{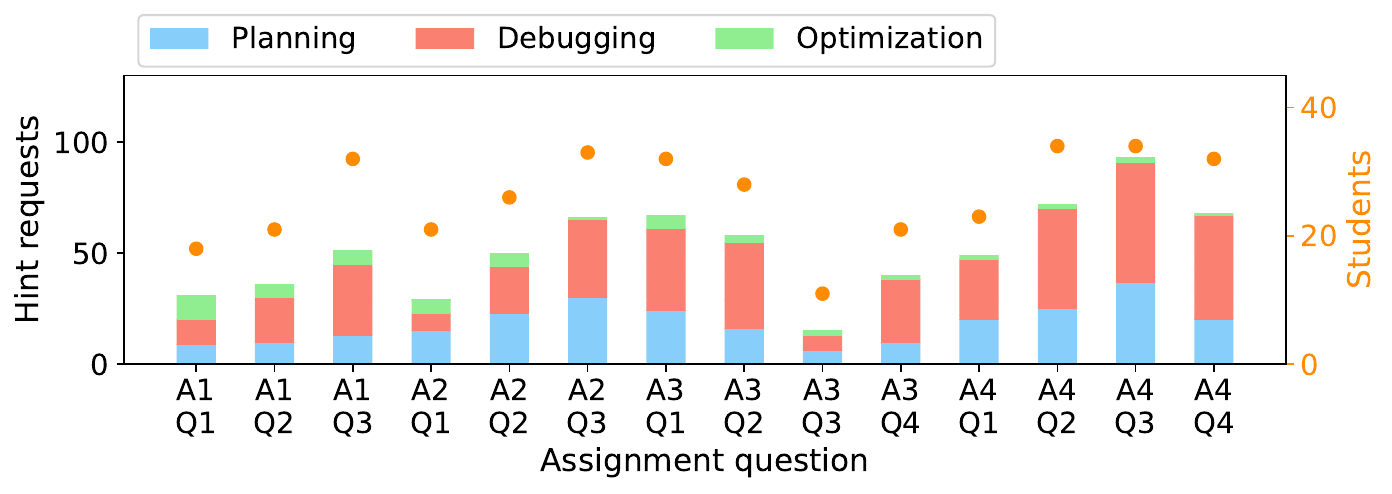}
    \label{fig:basic_stats.hint_requests_per_question}
    \vspace{-7mm}
    \caption{
         Overview of students and hint requests. The orange dots and the right y-axis indicate the number of students who requested at least one hint (of any type) for each question. The stacked bars and the left y-axis represent the total number of hint requests per question, categorized by AIMS hint types. In total, students requested 258 planning, 411 debugging, and 56 optimization hints.
    }
    \label{fig:basic_stats}
    \vspace{-2.5mm}
\end{figure}

%% file: 4_RQ1.tex
\section{RQ1: Impacts of AIMS Hints on Help-Seeking Behavior}  \label{sec:rq1}

This section addresses RQ1 by outlining our analysis setup (Section~\ref{sec:rq1.setup}), presenting results (Section~\ref{sec:rq1.results}), and discussing key findings (Section~\ref{sec:rq1.discussions}).

\subsection{Analysis Setup}  \label{sec:rq1.setup}
To evaluate the impact of AIMS hints on student help-seeking behavior, we decompose RQ1 into three sub-questions: [RQ1a]: How do students engage with and perceive AIMS hints?, [RQ1b]: What behavioral patterns emerge in students' interactions with AIMS hints?, and [RQ1c]: How do these patterns vary based on question difficulty and student competency?

\looseness-1For RQ1a, engagement is measured by contemplation time and hint revisits, while perception is assessed through students' hint ratings. The contemplation time is defined as the time between receiving a hint and performing the next major action (i.e., requesting another hint or submitting a solution). To exclude irrelevant delays such as students taking a break, only durations up to $t = 1$ hour are considered for analysis.\footnote{We note that other choices, such as $t = 0.5$ or $t = 2$ hours also yield similar results.} Hint revisits are quantified by counting the number of times a student expands hint widgets to view previous hints.

\looseness-1For RQ1b, we analyze the sequence of requested hint types, the frequency of a type being present in the sequence, the first-requested type, and the most-requested type. Each of these is counted based on all student-question pairs.

\looseness-1For RQ1c, we examine how hint sequences and the presence of hint types vary across question difficulty and student competency (as defined in Section~\ref{sec:study_setup.difficulty_competency}).

\subsection{Results}  \label{sec:rq1.results}

\looseness-1
\textbf{Engagement and perception of AIMS hints.} Figure~\ref{fig:result_behavior_engagement} summarizes engagement and perception results. A Mood's median test~\cite{mood1950introduction} reveals a significant difference in contemplation time across hint types ($p = 0.002$). Post hoc pairwise comparisons with Bonferroni correction~\cite{bonferroni1936teoria} confirm that the contemplation time on planning hints (median = $14.0$ minutes) is longer than debugging hints (median = $7.1$ minutes, $p = 0.006$). For hint revisits, a Kruskal-Wallis H test~\cite{kruskal1952use} indicates a significant difference ($p = 0.009$), with Dunn's post hoc tests using Bonferroni correction confirming more revisits for planning than for debugging hints ($p = 0.015$). Similarly, a $\chi^2$ test~\cite{Pearson1900} detects a significant difference in hint rating ($p = 0.003$), with pair-wise comparison using a Bonferroni correction confirming higher ratings for planning than for debugging hints ($p = 0.005$).
Optimization hints, with smaller sample sizes ($33$--$56$), do not show any significant differences.

\input{figs/result_behavior_engagement/fig_main}

\input{figs/result_behavior_patterns/fig_main}
\input{figs/result_behavior_comparison/fig_main}

\looseness-1\textbf{Help-seeking patterns.} Figure~\ref{fig:result_behavior_patterns} shows students' help-seeking patterns. Nine out of ten most frequent hint sequences consist of a single hint type, with debugging hints being requested the most, followed by planning hints, while optimization hints were rarely used.
When both planning and debugging hints were sought, planning hints were more likely to be requested first, aligning with metacognitive phases~\cite{schraw1995metacognitive}.
Notably, $43\%$ of optimization hints ($24$ out of $56$) were requested in isolation. 
Upon investigation of students' code and reflections, these cases often belonged to high-performing students who solved without hints and then sought further improvements. However, some other students seemed to mistakenly request these optimization hints when they needed debugging support.

\textbf{Behavioral patterns by difficulty and competency.} Figure~\ref{fig:result_behavior_comparison} compares help-seeking behaviors based on question difficulty (Figures~\ref{fig:result_behavior_comparison.easy} and~\ref{fig:result_behavior_comparison.hard}) and student competency (Figure~\ref{fig:result_behavior_comparison.weak} and~\ref{fig:result_behavior_comparison.strong}). As question difficulty increases, students request more debugging hints, while planning hint usage remains constant. In contrast, optimization hints were requested more often for easier questions (in $20$ student-question pairs) than harder ones ($10$ pairs).
Regarding student competency, higher-competency students request more hints overall, particularly planning hints, compared to their lower-competency peers.

\subsection{Discussions}  \label{sec:rq1.discussions}
\looseness-1Our findings reveal key patterns in students’ help-seeking behavior and hint usage. The high engagement and positive perception of planning hints suggest that structured guidance at the planning stage can be highly beneficial. However, it remains overlooked in many existing feedback systems, which primarily focus on debugging support~\cite{pankiewicz2024navigating,DBLP:conf/lak/PhungPS0CGSS24}. By emphasizing planning, educators and AI systems could better scaffold students’ problem-solving processes and reduce inefficient trial-and-error cycles. Despite the high perceived value of planning hints, students requested debugging hints more often, especially for harder questions. This suggests reactive rather than proactive strategies~\cite{alasmari2024teachers,hoffman2021students,DBLP:journals/frai/MallikG23}, where students rely on troubleshooting rather than strategic planning. While debugging is an essential skill, over-reliance on it may hinder deeper conceptual understanding. Future AI-assisted learning environments should promote proactive planning, encouraging students to articulate their problem-solving strategies before coding. Optimization hints, which were aimed at mastering skills rather than improving grades, were underutilized ($8\%$ of total requests). Further research is needed to make them more attractive and effective, fostering student mastery learning~\cite{bloom1968learning}.

Students predominantly requested a single hint type per question, indicating a potential lack of metacognitive awareness--they may not always recognize when they need planning, monitoring, or evaluation support~\cite{loksa2022metacognition,stanton2021fostering}. The variation in hint-seeking behavior by question difficulty further reinforces this issue. While debugging hints were requested more for harder questions, planning hint usage remained unchanged, even though structured planning is particularly useful for complex tasks~\cite{eichmann2019role}. Intelligent tutoring systems could use adaptive prompts or reflective exercises to help students better assess their difficulties~\cite{alzaid2018effectiveness,Choi2023-fg}. Additionally, designing interventions that make planning more explicit--such as requiring students to draft pseudocode before coding--could help bridge the gap.

The difference in hint usage between higher- and lower-competency students may provide further insights into how metacognition contributes to learning. While some studies reported that weaker students require more help~\cite{beal2010evaluation,beal2007line}, our results show that higher-competency students requested more hints overall, especially planning hints. This may be because they are more persistent in problem-solving and thus, are more willing to engage with available support. In contrast, lower-competency peers may be more prone to give up earlier. This difference likely contributes to performance disparities, as discussed next.

%% file: figs/result_behavior_engagement/fig_main.tex
\begin{figure}[t!]
    \centering
        \begin{subfigure}[b]{0.325\linewidth}
            \centering
            \scalebox{1.0}{
                \includegraphics[width=\linewidth]{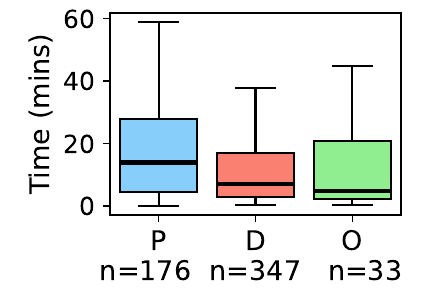}
            }
            \vspace{-7mm}
            \subcaption{Contemplation time}
            \label{fig:result_behavior_engagement.contemplation}
        \end{subfigure}
        \hfill
        \begin{subfigure}[b]{0.325\linewidth}
            \centering
            \scalebox{1.0}{
                \includegraphics[width=\linewidth]{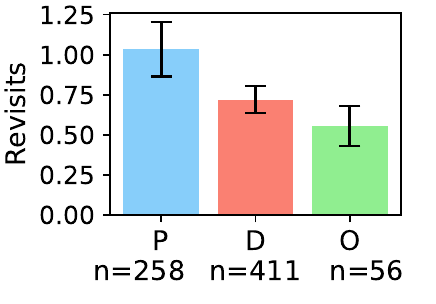}
            }
            \vspace{-7mm}
            \subcaption{Hint revisits}
            \label{fig:result_behavior_engagement.revisits}
        \end{subfigure}
        \hfill
        \begin{subfigure}[b]{0.325\linewidth}
            \centering
            \scalebox{1.0}{
                \includegraphics[width=\linewidth]{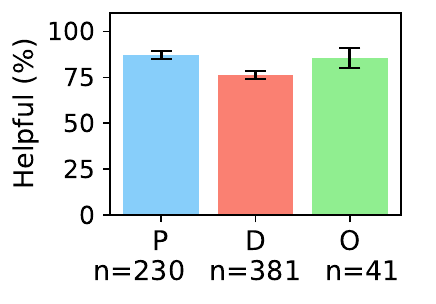}
            }
            \vspace{-7mm}
            \subcaption{Hint rating}
            \label{fig:result_behavior_engagement.rating}
        \end{subfigure}
    \vspace{-2mm}
    \caption{
        \looseness-1Results for RQ1a: Student engagement and perception of hints. (a) demonstrates the amount of time students contemplated after receiving a hint. (b) shows the number of revisits per hint. (c) presents the average rating of hints.
    }
    \label{fig:result_behavior_engagement}
    \vspace{-2mm}
\end{figure}

%% file: figs/result_behavior_patterns/fig_main.tex
\begin{figure}[t!]
    \centering
    \begin{subfigure}[b]{\linewidth}
        \centering
        \scalebox{1.0}{
            \includegraphics[width=\linewidth]{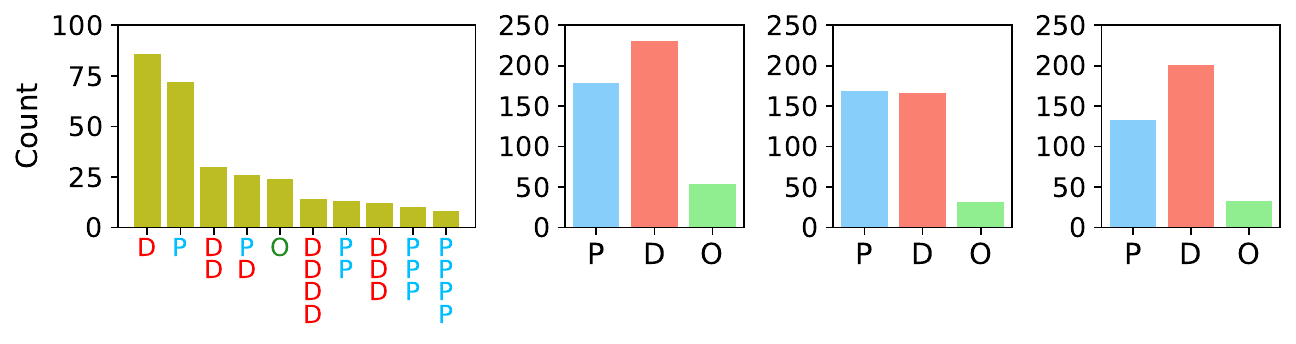}
        }
        \vspace{-6mm}
    \end{subfigure}
    \begin{subfigure}[b]{0.37\linewidth}
        \centering
        \vspace{-6mm}
        \subcaption{Hint sequence}
        \label{fig:result_behavior_patterns.sequence}
    \end{subfigure}
    %
    %
    \begin{subfigure}[b]{0.2\linewidth}
        \centering
        \vspace{-6mm}
        \subcaption{Type present}
        \label{fig:result_behavior_patterns.present}
    \end{subfigure}
    %
    %
    \begin{subfigure}[b]{0.2\linewidth}
        \centering
        \vspace{-6mm}
        \subcaption{First type}
        \label{fig:result_behavior_patterns.first}
    \end{subfigure}
    %
    %
    \begin{subfigure}[b]{0.205\linewidth}
        \centering
        \vspace{-6mm}
        \subcaption{Majority type}
        \label{fig:result_behavior_patterns.major}
    \end{subfigure}
    \vspace{-2mm}
    \caption{
        \looseness-1Results for RQ1b: Patterns of student hint usage. In all plots, y-axis represents the counts of student-question pairs. (a) displays the most common hint sequences; (b), (c), and (d) show the number of times a hint type is present in a sequence, is the first in a sequence, and is the majority in a sequence, respectively.
    }
    \label{fig:result_behavior_patterns}
    \vspace{-3.5mm}
\end{figure}

%% file: figs/result_behavior_comparison/fig_main.tex
\begin{figure}[t!]
    \centering
    \begin{subfigure}{0.48\linewidth}
        \centering
        \scalebox{1.0}{
            \includegraphics[width=\linewidth]{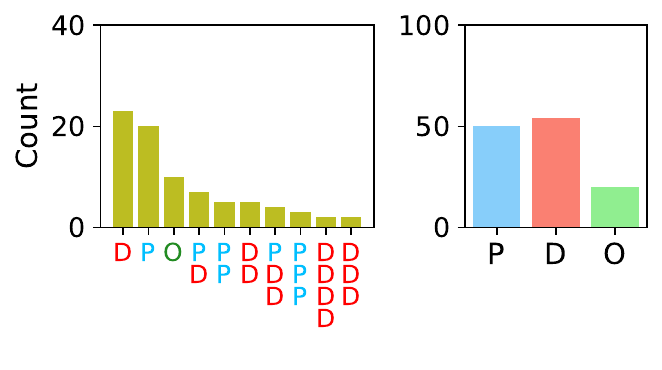}
        }
        \vspace{-8mm}
        \subcaption{Easier questions: 181 hints.}
        \label{fig:result_behavior_comparison.easy}
    \end{subfigure}
    \hfill
    \begin{subfigure}{0.48\linewidth}
        \centering
        \scalebox{1.0}{
            \includegraphics[width=\linewidth]{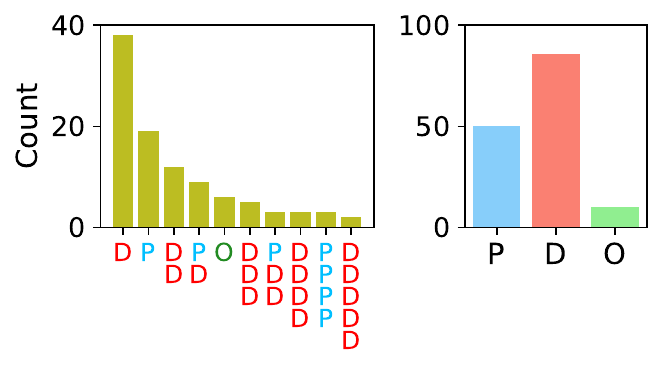}
        }
        \vspace{-8mm}
        \subcaption{Harder questions: 225 hints.}
        \label{fig:result_behavior_comparison.hard}
    \end{subfigure}
    \begin{subfigure}{0.48\linewidth}
        \centering
        \scalebox{1.0}{
            \includegraphics[width=\linewidth]{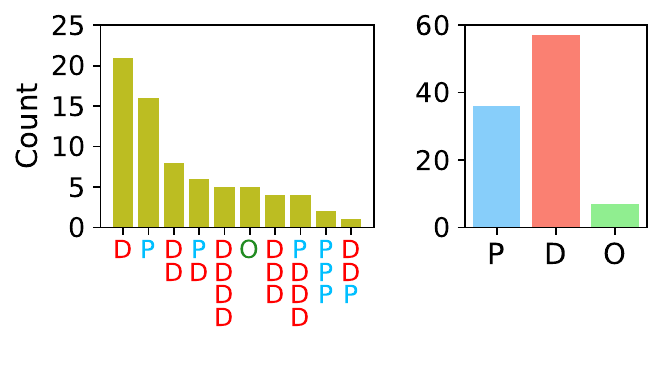}
        }
        \vspace{-10mm}
        \subcaption{Lower-competency students: 160 hints.}
        \label{fig:result_behavior_comparison.weak}
    \end{subfigure}
    \hfill
    \begin{subfigure}{0.48\linewidth}
        \centering
        \scalebox{1.0}{
            \includegraphics[width=\linewidth]{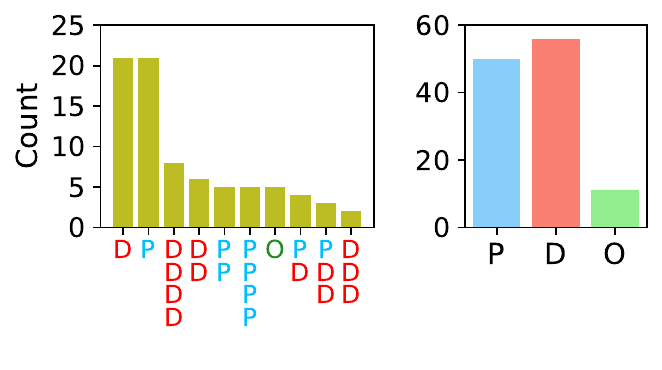}
        }
        \vspace{-10mm}
        \subcaption{Higher-competency students: 200 hints.}
        \label{fig:result_behavior_comparison.strong}
    \end{subfigure}
    \vspace{-2mm}
    \caption{
        Results for RQ1c: Behaviors by difficulty and competency. In each subfigure, the left shows Hint sequence and the right shows Type present. (a) and (b) present results by difficulty while (c) and (d) present results by competency.
    }
    \label{fig:result_behavior_comparison}
    \vspace{-5mm}
\end{figure}

%% file: 5_RQ2.tex
\section{RQ2: AIMS Hints and Performance}
Building on the behavioral patterns identified in RQ1, this section addresses RQ2: how those patterns relate to student problem-solving performance.

\subsection{Analysis Setup}
We break RQ2 into two sub-questions: [RQ2a]: How do students' interaction patterns with AIMS hints relate to their problem-solving performance? and [RQ2b]: How do these effects vary by question difficulty and student competency?
To answer these, we focus on students' final solving rates in relation to hint usage.

\subsection{Results}
\input{figs/result_performance/fig_main}

\textbf{Hint usage and overall performance.} As shown in Figure~\ref{fig:result_performance.all_all}, requesting planning hints is associated with significantly higher performance than no hints ($p=0.013$). Other types (debugging, optimization) show no significant effects.

\textbf{Performance variation by difficulty and competency.} Across all difficulty and competency conditions, planning hints are consistently (even though not always significantly) associated with higher performance than no hints (see 
 Figure~\ref{fig:result_performance}). 
In contrast, optimization hints are linked to lower performance on easier questions ($p=0.039$), particularly by lower-competency students ($p=0.003$). A closer examination of code and reflections reveals that in most of these cases, students misused optimization hints for debugging aid.
Only higher-competency students, but not lower-competency ones, exhibit significantly better performance associated with hint use.
Among higher-competency students, requesting any hints is linked to significantly higher performance ($p=0.007$), with independent positive effects for planning ($p=0.024$) and debugging hints ($p=0.044$).

\subsection{Discussions}
Our findings underscore a consistent association between planning hints and higher performance. This aligns with metacognitive theory, which emphasizes planning as a critical step in problem-solving~\cite{cohen1982planning,eichmann2019role,gunzelmann2003problem}. While the study did not establish causality (students' high intrinsic SRL skills could be a confounding factor that caused both requesting of planning hints and higher final performance), these results suggest potential instructional value from planning hints. In contrast, optimization hints are sometimes linked to lower performance, likely due to students' misuse, despite clear descriptions and availability of reference (see Section~\ref{sec:study_setup.hint_system}). This highlights the need for AI-assisted learning systems to ensure awareness of the available support and its alignment with their issues.

Our results differ from some studies that found hints more beneficial for lower-competency students~\cite{beal2010evaluation,beal2007line}. 
This may be because lower-competency students may have
weaker SRL skills, making it harder to leverage hints effectively. Additionally, since our hints are purposely Socratic and non-direct, lower-competency students may learn less from them than higher-competency ones.
Future research should explore varying hint detail levels to fit different student groups~\cite{mao2024effects,xiao2024exploring}.

%% file: figs/result_performance/fig_main.tex
\begin{figure}[t!]
    \centering
    \begin{minipage}{\linewidth}
        \centering
        \begin{subfigure}[b]{0.35\linewidth}
            \scalebox{1.0}{
                \includegraphics[height=0.125\paperheight]{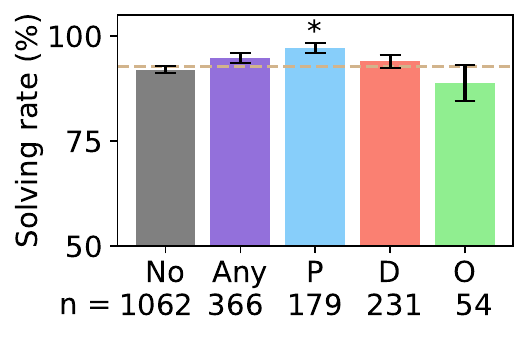}
            }
            \vspace{-6.5mm}  
            \subcaption{Stu: all, Ques: all}
            \label{fig:result_performance.all_all}
            \vspace{1mm}
        \end{subfigure}
        %
        %
        \begin{subfigure}[b]{0.31\linewidth}
            \hfill
            \scalebox{1.0}{
                \includegraphics[height=0.125\paperheight]{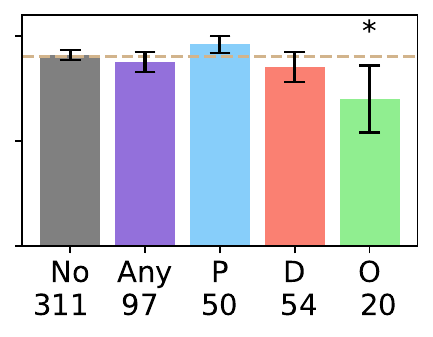}
            }
            \vspace{-2.3mm}
            \subcaption{Stu: all, Ques: easy}
            \label{fig:result_performance.easy_all}
            \vspace{1mm}
        \end{subfigure}
        \hfill
        \begin{subfigure}[b]{0.31\linewidth}
            \scalebox{1.0}{
                \includegraphics[height=0.125\paperheight]{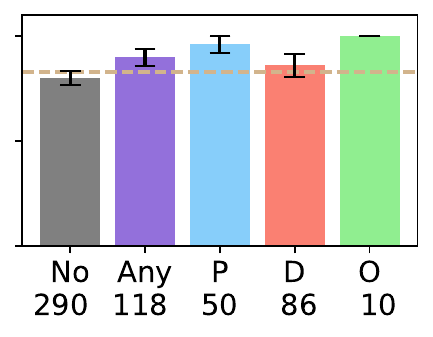}
            }
            \vspace{-2.3mm}
            \subcaption{Stu: all, Ques: hard}
            \label{fig:result_performance.hard_all}
            \vspace{1mm}
        \end{subfigure}
    \end{minipage}
    \begin{minipage}{\linewidth}
        \begin{subfigure}[b]{0.35\linewidth}
            \scalebox{1.0}{
                \includegraphics[height=0.125\paperheight]{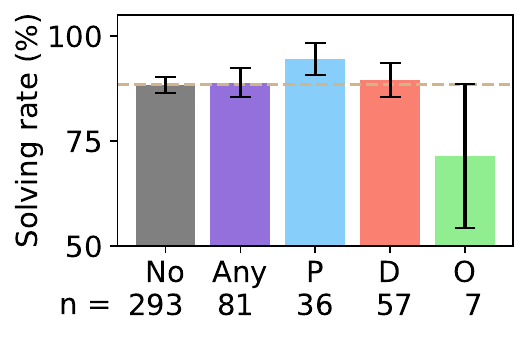}
            }
            \vspace{-6.5mm}
            \subcaption{Stu: lower, Ques: all}
            \label{fig:result_performance.all_weak}
            \vspace{1mm}
        \end{subfigure}
        %
        %
        \begin{subfigure}[b]{0.31\linewidth}
            \hfill
            \scalebox{1.0}{
                \includegraphics[height=0.125\paperheight]{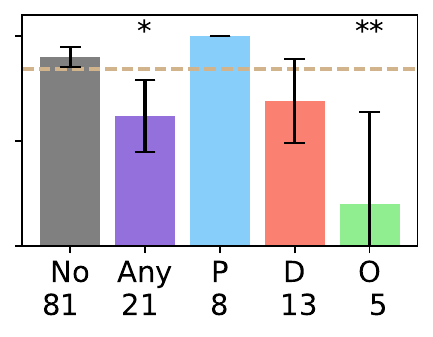}
            }
            \vspace{-2.3mm}
            \subcaption{Stu: lower, Ques: easy}
            \label{fig:result_performance.easy_weak}
            \vspace{1mm}
        \end{subfigure}
        \hfill
        \begin{subfigure}[b]{0.31\linewidth}
            \scalebox{1.0}{
                \includegraphics[height=0.125\paperheight]{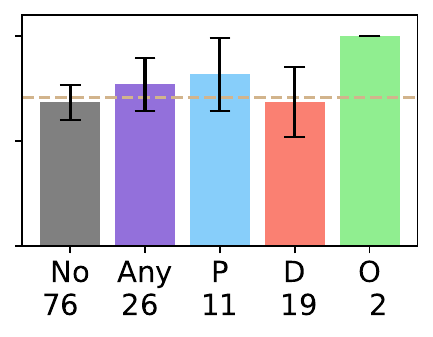}
            }
            \hfill
            \vspace{-2.3mm}
            \subcaption{Stu: lower, Ques: hard}
            \label{fig:result_performance.hard_weak}
            \vspace{1mm}
        \end{subfigure}
    \end{minipage}
    \begin{minipage}{\linewidth}
        \begin{subfigure}[b]{0.35\linewidth}
            \scalebox{1.0}{
                \includegraphics[height=0.125\paperheight]{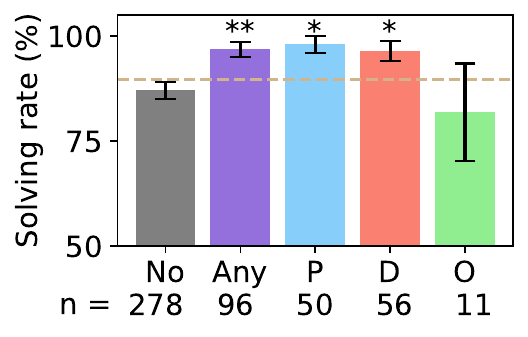}
            }
            \vspace{-6.5mm}
            \subcaption{Stu: higher, Ques: all}
            \label{fig:result_performance.all_strong}
        \end{subfigure}
        %
        %
        \begin{subfigure}[b]{0.31\linewidth}
            \hfill
            \scalebox{1.0}{
                \includegraphics[height=0.125\paperheight]{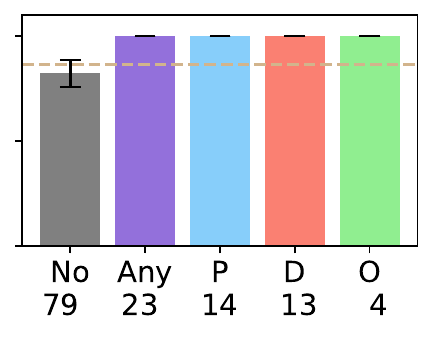}
            }
            \vspace{-2.3mm}
            \subcaption{Stu: higher, Ques: easy}
            \label{fig:result_performance.easy_strong}
        \end{subfigure}
        \hfill
        \begin{subfigure}[b]{0.31\linewidth}
            \scalebox{1.0}{
                \includegraphics[height=0.125\paperheight]{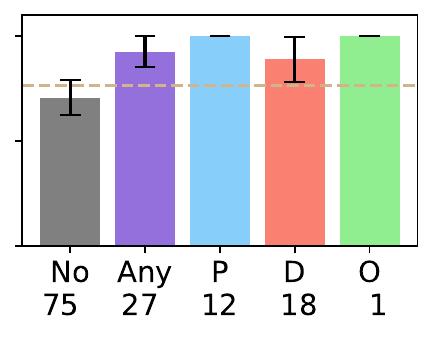}
            }
            \hfill
            \vspace{-2.3mm}
            \subcaption{Stu: higher, Ques: hard}
            \label{fig:result_performance.hard_strong}
        \end{subfigure}
    \end{minipage}
    \vspace{-1mm}
    \caption{
        Results for RQ2: Performance by help-seeking behavior. The dashed line depicts the average overall performance of the condition; the five bars represent no hint requested, any type requested, and each type present in the sequence of requested hints (i.e., Type present). The vertical lines indicate standard errors; \textbf{*} indicates a significant difference in performance to the `No' group w.r.t a $\chi^2$ test with $p<0.05$, while \textbf{**} indicates $p<0.01$. Y-axes are plotted from $50\%$.
    }
    \label{fig:result_performance}
    \vspace{-3mm}
\end{figure}

%% file: 6_limitations.tex
\section{Limitations}  \label{sec:limitations}
\looseness-1 This work has several limitations. First, it was conducted in a single programming course, where students could make multiple submissions, resulting in high overall grades ($>92\%$). This makes it challenging to isolate the impacts of AIMS hints on student performance. Future work should examine AIMS hints in diverse courses with varying grading schemes. 
Second, students' under-utilization of optimization hints limited our ability to assess their impact on students. However, this highlights an opportunity for future work on strategies to encourage students to pursue mastery learning beyond correct solutions. 
Third, we did not measure long-term learning gains. Future research should evaluate the long-term effects using methods such as retention tests, delayed post-tests, or longitudinal tracking. 
Fourth, we focused solely on button-based hints. Future work could explore alternative interfaces, such as chatbots or voice assistants. 
Finally, we investigated AIMS hints in isolation from other forms of support. Future studies should investigate their integration with complementary support, such as instructor-led office hours, to create a more comprehensive learning environment.

%% file: 7_conclusions.tex
\section{Conclusions}
\looseness-1This paper investigates the integration of metacognitive theory in AI-assisted programming education through a hint system aligned with planning, monitoring, and evaluation phases. By designing three corresponding hint types--planning, debugging, and optimization--and allowing students to select hints within a quota, our approach not only tailors the support to students' problem-solving stage but also fosters students' metacognitive awareness. A field study reveals that students engage most with planning hints, which are consistently linked to higher performance. However, students often request only one hint type per question and, when facing harder tasks, request more debugging but not more planning hints. This insight warrants future work on better metacognitive guidance in student awareness. Our findings provide empirical evidence of the synergy between AI and pedagogical theory in programming education, opening avenues for future research on pedagogically informed AI-tutoring systems.

%% file: 8_acknowledgements.tex

\textbf{Acknowledgments.} 
Funded/Co-funded by the European Union (ERC, TOPS, 101039090). Views and opinions expressed are however those of the author(s) only and do not necessarily reflect those of the European Union or the European Research Council. Neither the European Union nor the granting authority can be held responsible for them.